\documentclass[conference,10pt,letter]{IEEEtran}
\usepackage{graphicx}
\usepackage{amsmath}
\usepackage{amssymb}
\usepackage{calc}
\usepackage{multirow}
\usepackage{mathrsfs}
\usepackage[english]{babel}
\usepackage{float}
\usepackage{cool}
\usepackage{url}
\usepackage{nicefrac, xfrac}
\IEEEoverridecommandlockouts

\usepackage{times}
\usepackage{url}
\usepackage{color,colortbl}
\usepackage{tikz}
\usepackage{caption}
\usepackage{subcaption}
\usetikzlibrary{calc,arrows,positioning}
\definecolor{TableCol}{rgb}{1,0.75,0}
\usepackage{enumitem}
\usepackage{commath}
\usepackage{tensor}
\usepackage{mathtools}
\usepackage{amsmath,amsfonts,amssymb,commath}
\usepackage{gensymb}

\usepackage{acro}
\usepackage{siunitx}
\DeclareAcronym{adev}{
    short = ADEV,
    long = Allan deviation
}
\DeclareAcronym{avar}{
    short = AVAR,
    long = Allan variance
}
\DeclareAcronym{kf}{
    short = KF,
    long = Kalman filter
}
\DeclareAcronym{pdf}{
    short = PDF,
    long = probability density function
}
\DeclareAcronym{wp}{
    short = WP,
    long = work package
}
\DeclareAcronym{kpw}{
    short = KPW,
    long = Kalman filter plus weights
}
\DeclareAcronym{ai}{
    short = AI,
    long = artificial intelligence
}
\DeclareAcronym{fs}{
    short = FS,
    long = frequency standard
}
\DeclareAcronym{mmse}{
    short = MMSE,
    long = minimum mean square error
}
\DeclareAcronym{ism}{
    short = ISM,
    long = innovation sequence monitoring
}
\DeclareAcronym{solsep}{
    short = SolSep,
    long = solution separation
}
\DeclareAcronym{fd}{
    short = FD,
    long = fault detection
}
\DeclareAcronym{dof}{
    short = DoF,
    long = degree of freedom
}
\DeclareAcronym{fa}{
    short = FA,
    long = false alarm
}
\DeclareAcronym{md}{
    short = MD,
    long = missed detection
}
\DeclareAcronym{btse}{
    short = BTSE,
    long = basic time scale equation
}
\DeclareAcronym{tsu}{
    short = TSU,
    long = time source unit
}
\DeclareAcronym{pe}{
    short = PE,
    long = parity equations
}
\DeclareAcronym{rmse}{
    short = RMSE,
    long = root mean square error
}
\DeclareAcronym{fir}{
    short = FIR,
    long = finite impulse response
}
\DeclareAcronym{std}{
    short = STD,
    long = standard deviation
}
\DeclareAcronym{cusum}{
    short = CUSUM,
    long = cumulative sum
}
\DeclareAcronym{glrt}{
    short = GLRT,
    long = generalised likelihood ratio test
}
\DeclareAcronym{pv}{
    short = PV,
    long = position and velocity
}
\DeclareAcronym{blkavg}{
    short = BLKAVG,
    long = block averages
}

\DeclareAcronym{seqavg}{
    short = SEQAVG,
    long = sequential averages
}

\DeclareAcronym{gm}{
    short = GM,
    long = Gauss-Markov
}

\DeclareAcronym{nfe}{
    short = NFE,
    long = normalized frequency error
}

\newcommand{\bfa}{\ensuremath{\mathbf{a}}}

\newcommand{\bfe}{\ensuremath{\mathbf{e}}}
\newcommand{\bff}{\ensuremath{\mathbf{f}}}
\newcommand{\bfg}{\ensuremath{\mathbf{g}}}
\newcommand{\bfh}{\ensuremath{\mathbf{h}}}

\newcommand{\bfr}{\ensuremath{\mathbf{r}}}

\newcommand{\bfv}{\ensuremath{\mathbf{v}}}
\newcommand{\bfw}{\ensuremath{\mathbf{w}}}
\newcommand{\bfx}{\ensuremath{\mathbf{x}}}
\newcommand{\bfy}{\ensuremath{\mathbf{y}}}
\newcommand{\bfz}{\ensuremath{\mathbf{z}}}

\newcommand{\bfC}{\ensuremath{\mathbf{C}}}
\newcommand{\bfD}{\ensuremath{\mathbf{D}}}

\newcommand{\bfG}{\ensuremath{\mathbf{G}}}

\newcommand{\bfI}{\ensuremath{\mathbf{I}}}

\newcommand{\bfK}{\ensuremath{\mathbf{K}}}
\newcommand{\bfL}{\ensuremath{\mathbf{L}}}
\newcommand{\bfM}{\ensuremath{\mathbf{M}}}

\newcommand{\bfP}{\ensuremath{\mathbf{P}}}

\newcommand{\bfQ}{\ensuremath{\mathbf{Q}}}
\newcommand{\bfR}{\ensuremath{\mathbf{R}}}
\newcommand{\bfS}{\ensuremath{\mathbf{S}}}

\newcommand{\bbfx}{\ensuremath{\bar{\mathbf{x}}}}

\newcommand{\hbfx}{\ensuremath{\hat{\mathbf{x}}}}

\newcommand{\hbfz}{\ensuremath{\hat{\mathbf{z}}}}

\DeclareMathOperator*{\mean}{E}				
\DeclareMathOperator*{\tr}{tr}				

\newcommand{\bfxi}{\ensuremath{\boldsymbol{\xi}}}

\RequirePackage{xspace}

\usepackage{times}
\usepackage{amssymb}
\usepackage{url}

\usepackage{cite} 
\usepackage{stfloats}  

\newcounter{steps}
\newenvironment{descriptionA}
{\begin{list}{\textbf{Step} \textbf{\arabic{steps}}:}%
		{\setlength\labelsep{5pt}%
			\setlength\itemindent{5pt}%
			\setlength\parsep{0pt}%
			\setlength\leftmargin{0pt}%
			\setlength\labelwidth{0pt}%
			\usecounter{steps}}}%
	{\end{list}}

\def\real{\mathbb{R}}

\def\bfC{\mathbf C}
\def\bfD{\mathbf D}

\def\bfG{\mathbf G}

\def\bfI{\mathbf I}

\def\bfK{\mathbf K}
\def\bfL{\mathbf L}
\def\bfM{\mathbf M}

\def\bfQ{\mathbf Q}
\def\bfP{\mathbf P}
\def\bfR{\mathbf R}
\def\bfS{\mathbf S}

\def\bfa{\mathbf a}

\def\bfe{\mathbf e}
\def\bff{\mathbf f}
\def\bfg{\mathbf g}
\def\bfh{\mathbf h}

\def\bfv{\mathbf v}
\def\bfw{\mathbf w}
\def\bfx{\mathbf x}
\def\bfy{\mathbf y}
\def\bfz{\mathbf z}
\def\bfnul{\mathbf 0}

\def\calN{\mathcal{N}}

\def\mean{\mathsf{E}}

\def\bfnul{\mathbf 0}

\def\hbfx{\hat{\bfx}}
\def\hbfz{\hat{\bfz}}

\def\bbfx{\bar{\bfx}}

\def\tbfS{\tilde{\bfS}}

\def\bfxi{\ensuremath{\boldsymbol{\xi}}}

\begin{document}
	
	\onecolumn
	\vspace*{5cm}
	This paper has been accepted for publication in 2024 27th International Conference on Information Fusion (FUSION).
	Please cite the paper as:
	J. Duník, J. Matoušek, O. Straka, E. Blasch, J. Hiles and R. Niu, "Stochastic Integration Based Estimator: Robust Design and Stone Soup Implementation," 2024 27th International Conference on Information Fusion (FUSION), Venice, Italy, 2024, pp. 1-8, doi: 10.23919/FUSION59988.2024.10706476.
	\clearpage
	\twocolumn
	\title{Design of Unitless Normalized Measure of Nonlinearity for State Estimation}
	
\title{Stochastic Integration Based Estimator: Robust Design and Stone Soup Implementation}
	
\author{\IEEEauthorblockN{J. Duník, J. Matoušek, O. Straka}
	\IEEEauthorblockA{Dept. of Cyber., 
		Univ. of West Bohemia\\
		Univerzitn\'{i} 8, 306 14 Pilsen, Czech Rep.\\ Email: \{dunikj, matoujak, straka30\}@kky.zcu.cz}
	\and
	\IEEEauthorblockN{Erik Blasch}
	\IEEEauthorblockA{MOVEJ Analytics\\
      Fairborn, OH USA\\
	erik.blasch@gmail.com}
    \and
    \IEEEauthorblockN{John Hiles, Ruixin Niu}
	\IEEEauthorblockA{Dept. of ECE\\ Virginia Commonwealth University\\
	Email: \{hilesj, rniu\}@vcu.edu}
}

	\maketitle
	
	\selectlanguage{english}
	\begin{abstract}
	\noindent This paper deals with state estimation of nonlinear stochastic dynamic models. In particular, the stochastic integration rule, which provides asymptotically unbiased estimates of the moments of nonlinearly transformed Gaussian random variables, is reviewed together with the recently introduced stochastic integration filter (SIF). Using SIF, the respective multi-step prediction and smoothing algorithms are developed in \textit{full} and efficient \textit{square-root} form. The stochastic-integration-rule-based algorithms are implemented in Python (within the Stone Soup framework) and in MATLAB\textregistered\ and are numerically evaluated and compared with the well-known unscented and extended Kalman filters using the Stone Soup defined tracking scenario.
	\end{abstract}

	
\section{Introduction}
State estimation of discrete-time stochastic dynamic systems from noisy or incomplete measurements has been a subject of considerable research interest for the last decades. State estimation plays an important role in various fields such as navigation, speech and image processing, fault detection, optimal control, and tracking \cite{Yang:08}. 

Following the Bayesian approach, a general solution to the state estimation problem is given by the functional recursive relations (FRRs) for computing the probability density functions (PDFs) of the state conditioned on the measurements. The conditional PDFs provide a full description of the immeasurable state. The relations are, however, analytically tractable for a limited set of models where linearity is usually a common factor for exact Bayesian estimators, e.g., by the Kalman filter (KF) or the Rauch-Tung-Streibel smoother (RTSS). In other cases, an approximate solution to the FRRs has to be employed. These approximate estimation methods can be divided with respect to the validity of the estimates into global and local ones \cite{So:74,SiDu:09}.

\textit{Global} estimators, such as the particle or the point-mass approach, provide estimates in the form of conditional PDFs without any assumption on the conditional distribution family. These estimators are capable of estimating the state of a strongly nonlinear or non-Gaussian system but usually at the cost of higher computational demands. 

As opposed to global estimators, \textit{local}, or alternatively \textit{Gaussian}, estimators (GE) provide computationally efficient Gaussian-assumed estimates in the form of the conditional mean and covariance matrix only. Due to these simplifications, the GEs are especially suitable for mildly nonlinear models.

The GE can be further divided into the derivative and derivative-free estimators. Derivative estimators, developed in 70s, the nonlinear model is linearised using the Taylor expansion, which leads to the extended or second-order filter \cite{AnMo:79}. On the other hand, derivative-free estimators (DFEs) approximate the Gaussian-assumed conditional PDF by a set of weighted points which are propagated through the non-linear model. DFEs, being developed from 2000, are represented by the unscented, divided difference filter, or the cubature Kalman filter/smoother/predictor to name a few \cite{JuUhl:04,NoPoRa:00b,SiDu:09,DuStrSi:13}.

This paper focuses on the \textit{stochastic integration filter} (SIF) \cite{DuStrSi:13,DuStSiBl:15,KhReAb:17,ZhGu:17, StDu:17}, which compared to other Gaussian filters, provides an asymptotically exact integral evaluation needed for the conditional mean and covariance matrix calculation. As a consequence, the SIF was shown to provide superior estimation performance, still with acceptable computational complexity. Although, the SIF was thoroughly analysed and various extensions have been proposed (such as Student's t-distribution filter for heavy-tailed densities or $H_\infty$ filter for robust estimation), the areas of multi-step prediction, smoothing, and robust estimation have been left aside in the literature.

The goal of the paper is threefold: \textit{first}, to complete the family of stochastic integration estimators by the design of multi-step predictor and smoother, \textit{second}, to design efficient and robust versions of the estimators in the full and square-root forms, and \textit{third}, to implement the latest versions of the SIF and related predictor and smoother in the publicly available Python-based Stone Soup framework\footnote{\url{https://ostewg.isif.org/stone-soup/}}. In addition, the SIF and related predictor and smoother are  implemented also in MATLAB as a counterpart to the Python implementation.

The rest of the paper is organised as follows. In Section II, the state-space model and Bayesian estimation is introduced with the stress on a nonlinear Gaussian filter, smoother, and predictor design. Section III introduces the stochastic integration rule for calculation of moments of nonlinearly transformed Gaussian random variable and proposes the rule-based estimators. Then, square-root and other efficient formulations of the estimators are developed in Section IV and V. Subsequently, the Stone Soup framework and estimators' implementation are introduced in Section VI. Numerical illustration is provided in Section VII and the concluding remarks are drawn in Section VIII.

\section{State-space Model and Bayesian Estimation}
In this paper we consider a discrete-time nonlinear stochastic dynamic system, described by the state-space model
\begin{align}
    \bfx_{k+1}&=\bff_{k}(\bfx_{k})+\bfw_{k}, \label{eq:asx}\\
	\bfz_{k}&=\bfh_k(\bfx_k)+\bfv_k, \label{eq:asz}
\end{align}
where the vectors $\bfx_k\in\real^{n_x}$ and $\bfz_k\in\real^{n_z}$ represent the state of the system and the measurement at time instant $k$, respectively, and $k=0,1,2,\ldots,T$.  The state and measurement functions $\bff_k:\real^{n_x}\rightarrow\real^{n_x}$ and $\bfh_k:\real^{n_x}\rightarrow\real^{n_z}$ are supposed to be known. The state noise $\bfw_k\in\real^{n_x}$, the measurement noise $\bfv_k\in\real^{n_z}$, and the initial state $\bfx_0\in\real^{n_x}$ are supposed to be independent of each other.

The noises and the initial state are assumed to be normally distributed, i.e.,
\begin{align}
	p(\bfw_k)&=\calN\{\bfw_k;\bfnul_{n_x\times1},\bfQ_k\},\label{eq:pw}\\
	p(\bfv_k)&=\calN\{\bfv_k;\bfnul_{n_z\times1},\bfR_k\},\label{eq:pv}\\
	p(\bfx_0)&=\calN\{\bfx_0;\bbfx_0,\bfP_0\},\label{eq:px0}
\end{align}
where $\bfnul_{n_x\times1}\in\real^{n_x\times1}$ is a zero vector and the notation $\calN\{\bfx;\bbfx,\bfP\}$ stands for the Gaussian PDF of a random variable $\bfx$ with mean $\bbfx$ and covariance matrix $\bfP$. The first two moments of the random variables in \eqref{eq:pw}--\eqref{eq:px0} are supposed to be known.

\subsection{Bayesian State Estimation}
The FRRs are given by \cite{AnMo:79,SiDu:09}
\begin{align}
	p(\bfx_{k+m}|\bfz^k)&\!=\!\int\! p(\bfx_{k+m}|\bfx_{k+\ell}) p(\bfx_{k+\ell}|\bfz^k)d\bfx_{k+\ell},\label{eq:pred}\\
	p(\bfx_k|\bfz^k)&=\frac{p(\bfx_k|\bfz^{k-1})p(\bfz_k|\bfx_k)}{p(\bfz_k|\bfz^{k-1})},\label{eq:filt}\\
 p(\bfx_{k-m}|\bfz^k)&=p(\bfx_{k-m}|\bfz^{k-m})\int\frac{p(\bfx_{k-m+1}|\bfz^{k})}
         {p(\bfx_{k-\ell}|\bfz^{k-m})}\nonumber\\
                     &\times p(\bfx_{k-\ell}|\bfx_{k-m})d\bfx_{k-\ell}.\label{eq:smooth}
\end{align}
where $m>0, \ell=m-1$, and the symbol $\bfz^k$ represents the set of all measurements up to the time instant $k$, i.e., $\bfz^k=[\bfz_0,\bfz_1,\ldots\bfz_k]$. The PDF $p(\bfx_{k+m}|\bfz^{k-1})$ is the $m$-th step \textit{predictive} PDF computed by the Chapman-Kolmogorov equation \eqref{eq:pred}, $p(\bfx_k|\bfz^k)$ is the \textit{filtering} PDF computed by the Bayes' rule \eqref{eq:filt}, and $p(\bfx_{k-m}|\bfz^k)$ is the smoothed PDF calculated by \eqref{eq:smooth}. The PDFs $p(\bfx_{k}|\bfx_{k-1})$ and $p(\bfz_{k}|\bfx_{k})$ are the state transition PDF obtained from \eqref{eq:asx} and the measurement PDF obtained from \eqref{eq:asz}, respectively. The PDF $p(\bfz_k|\bfz^{k-1})=\int p(\bfx_k|\bfz^{k-1})p(\bfz_k|\bfx_k)d\bfx_k$ is the one-step predictive PDF of the measurement. The filtering and one-step predictive recursion \eqref{eq:pred}, \eqref{eq:filt} can be started from the initial PDF $p(\bfx_0|\bfz^{-1})=p(\bfx_0)$ and the recursion goes forward in time. The smoothing recursion \eqref{eq:smooth} starts with filtering PDF at the last time instant and the recursion goes backward in time.

Considering the nonlinear description of the system \eqref{eq:asx}--\eqref{eq:asz}, the FRRs are not exactly solvable. To allow the solution, the GEs assume the predictive conditional joint PDF, i.e.,
\begin{align}
	p(\bfx_{k+1},\bfz_{k+1}|\bfz^{k})\triangleq\calN\{\left[\begin{smallmatrix}\bfx_{k+1}\\\bfz_{k+1}\end{smallmatrix}\right]\!;\!\left[\begin{smallmatrix}\hbfx_{k+1|k}\\\hbfz_{k+1|k}\end{smallmatrix}\right]\!,\!\left[\begin{smallmatrix}\bfP^{xx}_{k+1|k} \bfP^{xz}_{k+1|k} \\ \bfP^{zx}_{k+1|k} \bfP^{zz}_{k+1|k}\end{smallmatrix}\right]\},\label{eq:jointPDFcond}
\end{align}
to be \textit{Gaussian}\footnote{This assumption is not always realistic especially for strongly nonlinear systems. However, the Gaussian PDF is fully defined by the first two moments, which allows efficient closed-form solution to the FRRs.}, which allows analytical (but inherently \textit{approximate}) solution to the FRRs leading to the following recursive GE 
algorithms for prediction, filtering, and smoothing\footnote{All GEs are linear algorithms with respect to the actual measurement and have the same structure as the KF and the RTSS.}, which form the general GE framework.

\subsection{Gaussian Predictor Design}
Assuming Gaussian conditional PDFs, the general solution to the multi-step prediction \eqref{eq:pred} in a GE framework is
\begin{align}
    p(\bfx_{k+m}|\bfz^k)\triangleq\calN\{\bfx_{k+m};\hbfx_{k+m|k},\bfP_{k+m|k}^{xx}\}
\end{align}
with the moments
\begin{align}
    \hbfx_{k+m|k}&=\mean[\bfx_{k+m}|\bfz^k]\label{eq:LPxpB}\\
    &=\!\int\!\bff_{k+\ell}(\bfx_{k+\ell})\calN\{\bfx_{k+\ell};\hbfx_{k+\ell|k}, \bfP^{xx}_{k+\ell|k}\}d\bfx_{k+\ell},\nonumber\\
    \bfP^{xx}_{k+m|k}&=\mean[(\bfx_{k+m}-\hbfx_{k+m|k})(\cdot)^T|\bfz^k]\nonumber\\
    &=\int(\bff_{k+\ell}(\bfx_{k+\ell})-\hbfx_{k+m|k})(\cdot)^T\nonumber\\
    &\times\calN\{\bfx_{k+\ell};\hbfx_{k+\ell|k}, \bfP^{xx}_{k+\ell|k}\}d\bfx_{k+\ell}+\bfQ_{k+\ell},\!\!\!\label{eq:LPPpB}
\end{align}
where the notation $(\bfa)(\cdot)^T$ means $(\bfa)(\bfa)^T$. It is clear that multi-step prediction can be deemed as recursive application of the one-step prediction.

\subsection{Gaussian Filter Design}
Calculation of the filtering PDF requires recursive evaluation of the Bayes' rule \eqref{eq:filt} and one-step predictive Chapman-Kolmogorov equation \eqref{eq:pred}. Assuming \eqref{eq:jointPDFcond}, their solution in a GE framework leads to Algorithm I:
\noindent\rule[-\baselineskip]{\linewidth}{1pt}
\vspace{-4pt}

\textbf{Algorithm I:} Generic Local Filter\\
\noindent\rule[\baselineskip]{\linewidth}{1pt}
\vspace{-1cm}
\begin{descriptionA}
	\item Set the time instant $k=0$ and define an 
	initial condition $p(\bfx_0|\bfz^{0})=\calN\{\bfx_0;\hbfx_{0|-1}, \bfP^{xx}_{0|-1}\}$.	 
	\item The moments of the filtering estimate $p(\bfx_{k}|\bfz^{k})\triangleq\calN\{\bfx_{k};\hbfx_{k|k}, \bfP^{xx}_{k|k}\}$ are 
	\begin{align}
		\hbfx_{k|k}&=\hbfx_{k|k-1}+\bfK_{k}(\bfz_{k}-\hbfz_{k|k-1})\enspace,\label{eq:LFxfB}\\
		\bfP^{xx}_{k|k}&=\bfP^{xx}_{k|k-1}-\bfK_{k}\bfP^{zz}_{k|k-1}\bfK_{k}^T\enspace,\label{eq:LFPfB}
	\end{align}
	where $\bfK_{k}=\bfP^{xz}_{k|k-1}(\bfP^{zz}_{k|k-1})^{-1}$ is the   
	filter gain,   
	\begin{align}
		\hbfz_{k|k-1}&\!\!=\!\!\int\!\!\bfh_{k}(\bfx_{k})\calN\{\bfx_{k};\hbfx_{k|k-1}, \bfP^{xx}_{k|k-1}\}d\bfx_{k},\label{eq:LFzpB}\\     
		\bfP^{zz}_{k|k-1}&=\int(\bfh_{k}(\bfx_{k})-\hbfz_{k|k-1})(\cdot)^T\nonumber\\
		&\times\calN\{\bfx_{k};\hbfx_{k|k-1}, \bfP^{xx}_{k|k-1}\}d\bfx_{k}+\bfR_{k}, \label{eq:LFPzB}\\
		\bfP^{xz}_{k|k-1}&=\int(\bfx_{k}-\hbfx_{k|k-1})(\bfh_{k}(\bfx_{k})-\hbfz_{k|k-1})^T\nonumber\\
		&\times\calN\{\bfx_{k};\hbfx_{k|k-1}, \bfP^{xx}_{k|k-1}\}d\bfx_{k}. \label{eq:LFPxzB}
	\end{align}
    \item The predictive moments of the Gaussian-assumed PDF $p(\bfx_{k+1}|\bfz^{k})\triangleq\calN\{\bfx_k;\hbfx_{k+1|k}, \bfP^{xx}_{k+1|k}\}$ are calculated according to \eqref{eq:LPxpB}, \eqref{eq:LPPpB} with $m=1$.
    
    The algorithm then continues to {\bf Step 2}.
\end{descriptionA}
\noindent\rule[\baselineskip]{\linewidth}{1pt}

\subsection{Gaussian Smoother Design}
The smoothing operation, i.e., solution to \eqref{eq:smooth}, can be generally divided into three categories:
\begin{itemize}
  \item {\it Fixed-point smoothing}, when time instant of the state to be estimated $k-m$ is fixed,
  \item {\it Fixed-lag smoothing}, when lag $m$ is fixed, and
  \item {\it Fixed-interval smoothing}, when time instant of the last available measurement $k$ is fixed.
\end{itemize}
Although each of these categories has its own dedicated smoothing algorithms, we will focus on the Rauch-Tung-Striebel-type smoother in this paper. This smoother can be easily adapted for all the categories \cite{SiDu:09} and has been used  for the vast majority of types of Gaussian filters, e.g., extended or unscented RTSS \cite{SiDu:06,SiDu:09,Sa:13}. 

The RTSS in the GE framework is given by the following relations for calculating the moments of the state
\begin{align}
    p(\bfx_{k-m}|\bfz^k)\triangleq\calN\{\bfx_{k-m};\hbfx_{k-m|k},\bfP_{k-m|k}^{xx}\}
\end{align}
as
\begin{align}
  \hbfx_{\mathfrak{m}|k}&=\hbfx_{\mathfrak{m}|\mathfrak{m}}+\bfL_{\mathfrak{m}}(\hbfx_{\mathfrak{m}+1|k}-\hbfx_{\mathfrak{m}+1|\mathfrak{m}}),\label{eq:rstsm_x}\\
  \bfP_{\mathfrak{m}|k}^{xx}&=\bfP_{\mathfrak{m}|\mathfrak{m}}^{xx}-\bfL_{\mathfrak{m}}(\bfP_{\mathfrak{m}+1|\mathfrak{m}}^{xx}-\bfP_{\mathfrak{m}+1|k}^{xx}) \bfL_{\mathfrak{m}}^T,\label{eq:rstsm_P}\\
  \bfL_{\mathfrak{m}}&=\bfP_{\mathfrak{m},\mathfrak{m}+1|\mathfrak{m}}^{xx}(\bfP_{\mathfrak{m}+1|\mathfrak{m}}^{xx})^{-1},\label{eq:rstsm_K}
\end{align}
where the lag $\mathfrak{m}=k-m$ and 
\begin{align}
    \bfP_{\mathfrak{m},\mathfrak{m}+1|\mathfrak{m}}^{xx}=\mean[(\bfx_\mathfrak{m}-\hbfx_{\mathfrak{m}|\mathfrak{m}})(\bfx_{\mathfrak{m}+1}-\hbfx_{\mathfrak{m}+1|\mathfrak{m}})^T|\bfz^\mathfrak{m}].\label{eq:rstsm_Pxx}
\end{align}
The RTSS is a backward recursion that processes the filtering and predictive estimates calculated by the filter and predictor in their forward run.

\section{Stochastic Integration Rule and Estimators}
Relations \eqref{eq:LPxpB}--\eqref{eq:rstsm_Pxx} represent a common framework for all the GE. The particular estimators differ in the calculation of the expected values 
\begin{itemize}
    \item \eqref{eq:LPxpB}, \eqref{eq:LPPpB} for prediction, 
    \item \eqref{eq:LFzpB}--\eqref{eq:LFPxzB} for filtering, and 
    \item \eqref{eq:rstsm_Pxx} for smoothing.
\end{itemize}

Calculation of the required moments lies in an evaluation  of the \textit{Gaussian-weighted} integral
\begin{align}
    \mathcal{I}=\mean[\bfg(\bfx_q)|\bfz^l]=\int \bfg(\bfx_q) \calN\{\bfx_q;\hbfx_{q|l},\bfP_{q|l}^{xx}\}d\bfx_q,\label{eq:int}
\end{align}
where the conditional mean $\hbfx_{q|l}$ and the covariance matrix $\bfP_{q|l}^{xx}$ are known from estimator preceding steps and the nonlinear function $\bfg(\cdot)$ stems from the model and calculated moment definition. The moment calculation can be, thus, seen as a calculation of the moments of a nonlinearly transformed Gaussian random variable with known description \cite{SiDu:09,NoPoRa:00b}. The particular form of $\bfg(\cdot)$, dimension of $\mathcal{I}$ denoted as $n_\mathcal{I}$, and the time indices $q,l$ depend on the selected estimator and are detailed later.

\subsection{Stochastic Integration Rule}
As any other integration rule (IR), the stochastic integration rule (SIR) \cite{GeMo:98,DuStrSi:13a,DuStSiBl:15} numerically evaluates \eqref{eq:int} using
\begin{align}
    \hat{\mathcal{I}}=\sum_{i=1}^S\omega^{(i)}\bfg(\bfxi_{k|l}^{(i)}),\label{eq:sir}
\end{align}
where $\bfxi_{q|l}^{(i)}$ and $\omega^{(i)}$ are spherical-radial points and corresponding weights of the rule, respectively. The SIR combines a deterministic IR and Monte-Carlo (MC) IR into a single algorithm, where some points  are selected randomly, and the remaining are computed using deterministic rules. The SIR takes the best properties of both, namely \cite{DuStSiBl:15}
\begin{itemize}
    \item Integration error going to zero with increasing $S$, i.e.
    \begin{align}
        \varepsilon=({\mathcal{I}}-\hat{\mathcal{I}})\rightarrow\bfnul\ \mathrm{as}\ S\rightarrow\infty,
    \end{align}
    and $\varepsilon=0$ if $\bfg(\cdot)$ is degree-$d$ polynomial, where $d$ is degree of used spherical and radial rule,
    \item Faster convergence rate than MC-based IRs,
    \item Implicit assessment of integral \eqref{eq:sir} evaluation error.
\end{itemize}
Full derivation of the SIR of degrees one, three, and five can be found in \cite{DuStSiBl:15}. In the algorithm below we briefly introduce 3rd-order SIR, which offers a reasonable compromise between accuracy and computational complexity.

\noindent\rule[-\baselineskip]{\linewidth}{1pt}
\vspace{-4pt}

\textbf{Algorithm II:} Degree 3 Stochastic Integration Rule\\
\noindent\rule[\baselineskip]{\linewidth}{1pt}
\vspace{-1cm}

\begin{descriptionA}
  \item Select  a maximum number of iterations  $N_{\max}$ or an error tolerance $\varepsilon_\mathrm{min}$.
  \item Set the current iteration number $N=0$, initial value of the integral 
    $\hat{\mathcal{I}}_{N}=\bfnul_{n_\mathcal{I}\times1}$, and the  initial square error 
    of the integral $\Sigma_{N}=\bfnul_{n_\mathcal{I}\times n_\mathcal{I}}$, and define the central IR point
    $\bfxi_{q|l}^{(0)}=\hbfx_{k|l}$, which is constant over all SIR iterations.
 \item Repeat (until $N=N_{max}$ or $\tr(\Sigma_N)<\varepsilon_\mathrm{min}$, but at least once) the 
   following loop:
  \begin{description}
    \item[a)] Set $N=N+1$.
    \item[b)] Generate a uniformly random orthogonal matrix $\bfC^{(N)}$ of dimension $n_x\times n_x$ \cite{Arn:04} and generate a random number $\rho$ from the \textit{Chi} distribution with $(n_x+2)$ degrees of freedom, i.e., $\rho^{(N)}\sim\mathrm{Chi}(n_x+2)$.
    \item[c)] Compute a set of IR points and weights according to
      \begin{align}
      \bfxi_{q|l}^{(N,i)}&=-\rho^{(N)}\bfS^{xx}_{q|l}\bfC^{(N)}\bfe_i+\hbfx_{q|l},\label{eq:chin3}\\ 
      \bfxi_{q|l}^{(N,n_x+i)}&=\rho^{(N)}\bfS^{xx}_{q|l}\bfC^{(N)}\bfe_i+\hbfx_{q|l}, \\
        \omega_{q|l}^{(N,0)}&=1-\tfrac{n_x}{\rho^2},\ \omega_{q|l}^{(N,i)}=\omega_{q|l}^{(N,n_x+i)}=\tfrac{1}{2\rho^2},\label{eq:chin5}
      \end{align}
      where $i=1,2,\ldots,n_x$, $\bfe_i$ is the $i$-th column of the identity matrix $\bfI_{n_x}$, and $\bfS^{xx}_{q|l}$ is the square-root of matrix $\bfP^{xx}_{q|l}$ so that $\bfP^{xx}_{q|l}=\bfS^{xx}_{q|l}\left(\bfS^{xx}_{q|l}\right)^T$. Matrix $\bfC^{(N)}$ rotates the IR points and $\rho^{(N)}$ scales them.
    \item[d)] Compute approximation of the integral 
      at current iteration $\mathfrak{I}^{(N)}$ (SR denotes a spherical-radial SIR), the update of the integral value 
      $\hat{\mathcal{I}}_{N}$, and the corresponding error $\Sigma_{N}$
      \begin{align}
       \mathfrak{I}^{(N)}&=\sum_{i=0}^{2n_x}\bfg(\bfxi_{q|l}^{(N,i)})\omega_{q|l}^{(N,i)},\label{eq:int_valSI_SIR3}\\
        \hat{\mathcal{I}}_N&=\hat{\mathcal{I}}_{N-1}+(\mathfrak{I}^{(N)}-\hat{\mathcal{I}}_{N-1})/N, \label{eq:int_val_SIR3}\\
        \Sigma_{N}&= (N-2)\Sigma_{N-1}/N+(\mathfrak{I}^{(N)}-\hat{\mathcal{I}}_{N-1})(\cdot)^T/N^2. \label{eq:int_sig_SIR3}
      \end{align}
  \end{description}
 \item Once a stopping condition is fulfilled, the calculated approximate value of the integral $\mathcal{I}$ \eqref{eq:int} is $\hat{\mathcal{I}}=\hat{\mathcal{I}}_{N}$ and total number of the IR points is $S=N(2n_x+1)$.
\end{descriptionA}
\noindent\rule[\baselineskip]{\linewidth}{1pt}
Note that the \textit{cubature} IR \cite{ArHa:09,SaOrAcAc:21}, widely used in the filter design, is a special case of the introduced SIR.

\subsection{SIR-based Prediction, Filtering, and Smoothing}
Design of the SIR-based predictor, filter, and predictor lies in calculation of the state and measurement moments in the GE framework using the SIR given in Algorithm II.
\subsubsection{Prediction} The predictive state mean $\hbfx_{k+m|k}$ \eqref{eq:LPxpB} is calculated using the SIR \eqref{eq:sir} detailed in Algorithm II, where $\hbfx_{q|l}=\hbfx_{k+\ell|k}$, $\bfP^{xx}_{q|l}=\bfP^{xx}_{k+\ell|k}$, and $\bfg(\bfx_q)=\bff_{k+\ell}(\bfx_{k+\ell})$ with $\ell=m-1$. The predictive covariance matrix is calculated analogously with the only change in the definition of the nonlinear function, which is $\bfg(\bfx_q)=\bigl(\bff_{k+\ell}(\bfx_{k+\ell})-\hbfx_{k+m|k}\bigr)\bigl(\bff_{k+\ell}(\bfx_{k+\ell})-\hbfx_{k+m|k}\bigr)^T$ in this case.
\subsubsection{Filtering} The predictive measurement mean $\hbfz_{k|k-1}$ \eqref{eq:LFPzB}, covariance matrix $\bfP^{zz}_{k|k-1}$ \eqref{eq:LFPzB}, and ``cross-covariance'' matrix $\bfP^{xz}_{k|k-1}$ \eqref{eq:LFPxzB} are calculated using the SIR \eqref{eq:sir}, where $\hbfx_{q|l}=\hbfx_{k|k-1}$, $\bfP^{xx}_{q|l}=\bfP^{xx}_{k|k-1}$, and the SIR-related nonlinear function is defined $\bfg(\bfx_q)=\bfh_{k}(\bfx_{k})$, $\bfg(\bfx_q)=\bigl(\bfh_{k}(\bfx_{k})-\hbfz_{k|k-1}\bigr)\bigl(\bfh_{k}(\bfx_{k})-\hbfz_{k|k-1}\bigr)^T$, and $\bfg(\bfx_q)=\bigl(\bfx_{k}-\hbfx_{k|k-1}\bigr)\bigl(\bfh_{k}(\bfx_{k})-\hbfz_{k|k-1}\bigr)^T$.
\subsubsection{Smoothing} The cross-covariance matrix $\bfP_{\mathfrak{m}+1,\mathfrak{m}|\mathfrak{m}}^{xx}$ \eqref{eq:rstsm_Pxx} for smoother gain is calculated using the SIR \eqref{eq:sir}, where $\hbfx_{q|l}=\hbfx_{\mathfrak{m}|\mathfrak{m}}$, $\bfP^{xx}_{q|l}=\bfP^{xx}_{\mathfrak{m}|\mathfrak{m}}$, and the SIR-related nonlinear function is $\bfg(\bfx_q)=\bigl(\bfx_\mathfrak{m}-\hbfx_{\mathfrak{m}|\mathfrak{m}}\bigr)\bigl(\bff_{\mathfrak{m}}(\bfx_\mathfrak{m})-\hbfx_{\mathfrak{m}+1|\mathfrak{m}}\bigr)^T$, respectively.

\section{Square-Root Estimators}
Numerical stability has been a focal point of Gaussian estimators development for the past several decades \cite{AnMo:79,Si:06,SiDu:09}. The emphasis has been placed on calculations that ensure symmetric and positive-definite conditional covariance matrix. Generally, two types of numerically robust algorithms have been developed
\begin{itemize}
    \item Filtering and smoothing covariance matrices \eqref{eq:LFPfB}, \eqref{eq:rstsm_P} calculation, where subtraction of two covariance matrices is omitted using the Joseph form~\cite{Si:06},
    \item Square-root implementation, where instead of the conditional covariance matrix, its factor is directly calculated.
\end{itemize}
The square-root implementation might lead to computational complexity reduction, as the covariance matrix factor $\bfS^{xx}_{q|l}$ required by the SIR in IR points calculation \eqref{eq:chin3} is directly calculated and thus available without the need of any decomposition.

In this section, we, therefore, propose the square-root version of the SIR-based estimators using \textit{direct calculation} of the state estimate covariance matrix factor for the prediction, filtering, and smoothing. Derivation of the square-root SIR-based estimators adopts the concept of the square-root unscented Kalman estimators developed in \cite{SiDu:05,SiDu:06}.

\subsection{SIR Reformulation}
The SIR points and weights $\{\bfxi_{q|l}^{(n,i)}, \omega_{q|l}^{(n,i)}\}_{i=0}^{2n_x}$ with $n=1,\ldots,N$, generated in $N$ iterations, can be concatenated into the following set of points and corresponding weights
\begin{align}
    \boldsymbol{\Xi}_{q|l}&=\{\bfxi_{q|l}^{(0)},\bfxi_{q|l}^{(1,1)}\!\!\!,\ldots,\bfxi_{q|l}^{(1,2n_x)}\!\!\!,\ldots,\bfxi_{q|l}^{(N,1)}\!\!\!,\ldots,\bfxi_{q|l}^{(N,2n_x)}\},\label{eq:allPoints}\\
    \boldsymbol{\Omega}_{q|l}&=\{\tilde{\omega}_{q|l}^{(0)},\tilde{\omega}_{q|l}^{(1,1)}\!\!\!,\ldots,\tilde{\omega}_{q|l}^{(1,2n_x)}\!\!\!,\ldots,\tilde{\omega}_{q|l}^{(N,1)}\!\!\!,\ldots,\tilde{\omega}_{q|l}^{(N,2n_x)}\},\label{eq:allWeights}
\end{align}
where $\tilde{\omega}_{q|l}^{(n,i)}=\nicefrac{\omega_{q|l}^{(n,i)}}{N}$ and $\tilde{\omega}_{q|l}^{(0)}=\tfrac{1}{N}\sum_{n=1}^N\omega_{q|l}^{(n,0)}$. Number of all SIR points is $n_S=2n_xN+1$, The proof is outlined in Appendix A.

For the sake of notional simplicity, we further denote $i$-th elements of sets $\boldsymbol{\Xi}_{k+\ell|k}, \boldsymbol{\Omega}_{k+\ell|k}$ \eqref{eq:allPoints}, \eqref{eq:allWeights} as $\boldsymbol{\Xi}_{k+\ell|k}^{(i)}, \boldsymbol{\Omega}_{k+\ell|k}^{(i)}$, respectively, where $i=0,1,\ldots,n_S$.

\subsection{Predictive Covariance Matrix Factor}
SIR-based predictive covariance $\bfP^{xx}_{k+m|k}$, stemming from \eqref{eq:LPPpB} and Section III.B.1, reads
\begin{align}
    \bfP^{xx}_{k+m|k}\!\! &=\!\! \sum^{n_S}_{i=1}
   \boldsymbol{\Omega}_{k+\ell|k}^{(i)}(\bff_{k+\ell}(\boldsymbol{\Xi}_{k+\ell|k}^{(i)})-\hat{\bfx}_{k+m|k})(\cdot)^T\!+\!\bfQ_{k+\ell}. \label{eq:predCov}
\end{align}

A factor of the covariance matrix $\bfP^{xx}_{k+m|k}$ \eqref{eq:predCov} is simply
\begin{align}
    \sqrt{\bfP^{xx}_{k+m|k}} = [\tilde{\bfS}^{xx}_{k+m|k},\ \bfS^Q_{k+\ell}], \label{eq:predFact}
\end{align}
where 
\begin{align}
    &\tilde{\bfS}^{xx}_{k+m|k} = 
   \Big[\sqrt{\boldsymbol{\Omega}_{k+\ell|k}^{(1)}}\left(\bff_{k+\ell}(\boldsymbol{\Xi}_{k+\ell|k}^{(i)})-\hat{\bfx}_{k+m|k}\right),\ldots,\nonumber\\
   &\sqrt{\boldsymbol{\Omega}_{k+\ell|k}^{(n_S)}}\left(\bff_{k+\ell}(\boldsymbol{\Xi}_{k+\ell|k}^{(i)})-\hat{\bfx}_{k+m|k}\right)\Big]\in\real^{n_x\times n_S}
\end{align}
and $\bfS^Q_{k+\ell}$ is a factor of the state noise covariance matrix s.t. $\bfQ_{k+\ell}=\bfS^Q_{k+\ell}(\bfS^Q_{k+\ell})^T$. Unfortunately, this factor is a \textit{rectangular} matrix unsuitable for further calculations. To make the matrix \textit{square}, we can apply the Householder triangularisation \cite{NoPoRa:00b,SiDu:05} denoted as $\mathrm{ht}(\cdot)$, i.e., 
\begin{align}
    \bfS^{xx}_{k+m|k} &= \mathrm{ht}(\tilde{\bfS}^{xx}_{k+m|k})\in\real^{n_x\times n_x}, \label{eq:predFactFin}
\end{align}
with $\bfP^{xx}_{k+m|k}=\tilde{\bfS}^{xx}_{k+m|k}(\tilde{\bfS}^{xx}_{k+m|k})^T=\bfS^{xx}_{k+m|k}(\bfS^{xx}_{k+m|k})^T$. Optionally, other decomposition, such as the QR decomposition, can be used as well for square matrix design.

\subsection{Filtering Covariance Matrix Factor}
Factorisation of the filtering covariance matrix $\bfP^{xx}_{k|k}$ is more involved due to subtraction of two terms on right-hand side of \eqref{eq:LFPfB}. Using a few matrix manipulations given in Appendix B, we can show that the filtering covariance matrix \eqref{eq:LFPfB} can be written as \cite{SiDu:05}
\begin{align}
    \bfP^{xx}_{k|k}=\left[\tbfS^{xx}_{k|k-1}-\bfK_k\tilde{\bfS}^{zz}_{k|k-1},\ \bfK_k\bfS^R_{k}\right][\cdot]^T,\label{eq:LFPfBFF}
\end{align}
where $\bfS^R_{k}$ is a factor of the measurement noise covariance matrix s.t. $\bfR_{k}=\bfS^R_{k}(\bfS^R_{k})^T$ and IR points stacking matrices are defined as
\begin{align}
    &{\tbfS}^{xx}_{k|k-1} = 
   \Big[\sqrt{\boldsymbol{\Omega}_{k|k-1}^{(1)}}\left(\boldsymbol{\Xi}_{k|k-1}^{(1)}-\hat{\bfx}_{k|k-1}\right),\ldots,\nonumber\\
   &\sqrt{\boldsymbol{\Omega}_{k|k-1}^{(n_S)}}\left(\boldsymbol{\Xi}_{k|k-1}^{(n_S)}-\hat{\bfx}_{k|k-1}\right)\Big]. \label{eq:predtSx}\\
    &\tilde{\bfS}^{zz}_{k|k-1} = 
   \Big[\sqrt{\boldsymbol{\Omega}_{k|k-1}^{(1)}}\left(\bfh_{k}(\boldsymbol{\Xi}_{k|k-1}^{(1)})-\hat{\bfz}_{k|k-1}\right),\ldots,\nonumber\\
   &\sqrt{\boldsymbol{\Omega}_{k|k-1}^{(n_S)}}\left(\bfh_{k}(\boldsymbol{\Xi}_{k|k-1}^{(n_S)})-\hat{\bfz}_{k|k-1}\right)\Big]. \label{eq:predtSzz}
\end{align}
Then, the matrix square factor can directly be determined as
\begin{align}
    \bfS^{xx}_{k|k}=\mathrm{ht}\left(\left[\tbfS^{xx}_{k|k-1}-\bfK_k\tilde{\bfS}^{zz}_{k|k-1},\ \bfK_k\bfS^R_{k}\right]\right).
\end{align}
In the square-root version, the Kalman gain reads
\begin{align}
    \bfK_k=\bfP^{xz}_{k|k-1}([\tbfS^{zz}_{k|k-1},\ \bfS^R_{k}][\cdot]^T)^{-1},
\end{align}
where $\bfP^{xz}_{k|k-1}={\tbfS}^{xx}_{k|k-1}(\tbfS^{zz}_{k|k-1})^T$.

\subsection{Smoothing Covariance Matrix Factor}
Similarly to the filtering matrix, smoothing covariance matrix $\bfP_{\mathfrak{m}|k}^{xx}$ \eqref{eq:rstsm_P} can be written in the form 
\begin{align}
    \bfP_{\mathfrak{m}|k}^{xx}=\left[\tbfS^{xx}_{\mathfrak{m}|\mathfrak{m}}-\bfL_\mathfrak{m}\tilde{\bfS}^{xx}_{\mathfrak{m}+1|\mathfrak{m}},\ \bfL_\mathfrak{m}\bfS^Q_{\mathfrak{m}},\ \bfL_\mathfrak{m}\bfS_{\mathfrak{m}+1|k}^{xx}\right][\cdot]^T,\label{eq:auxPs}
\end{align}
where IR points stacking matrices are defined as
\begin{align}
    &{\tbfS}^{xx}_{\mathfrak{m}|\mathfrak{m}} = 
   \Big[\sqrt{\boldsymbol{\Omega}_{\mathfrak{m}|\mathfrak{m}}^{(1)}}\left(\boldsymbol{\Xi}_{\mathfrak{m}|\mathfrak{m}}^{(1)}-\hat{\bfx}_{\mathfrak{m}|\mathfrak{m}}\right),\ldots,\nonumber\\
   &\sqrt{\boldsymbol{\Omega}_{\mathfrak{m}|\mathfrak{m}}^{(n_S)}}\left(\boldsymbol{\Xi}_{\mathfrak{m}|\mathfrak{m}}^{(n_S)}-\hat{\bfx}_{\mathfrak{m}|\mathfrak{m}}\right)\Big]. \label{eq:predtSxxSM}\\
    &\tilde{\bfS}^{xx}_{\mathfrak{m}+1|\mathfrak{m}} = 
   \Big[\sqrt{\boldsymbol{\Omega}_{\mathfrak{m}+1|\mathfrak{m}}^{(1)}}\left(\bff_{\mathfrak{m}}(\boldsymbol{\Xi}_{\mathfrak{m}+1|\mathfrak{m}}^{(1)})-\hat{\bfx}_{\mathfrak{m}+1|\mathfrak{m}}\right),\ldots,\nonumber\\
   &\sqrt{\boldsymbol{\Omega}_{\mathfrak{m}+1|\mathfrak{m}}^{(n_S)}}\left(\bff_{\mathfrak{m}}(\boldsymbol{\Xi}_{\mathfrak{m}+1|k\mathfrak{m}}^{(n_S)})-\hat{\bfx}_{\mathfrak{m}+1|\mathfrak{m}}\right)\Big]. \label{eq:predtSxxpSM}
\end{align}
Then, the the square factor of the covariance matrix \eqref{eq:auxPs} simply reads
\begin{align}
    \bfS_{\mathfrak{m}|k}^{xx}=\mathrm{ht}\left(\left[\tbfS^{xx}_{\mathfrak{m}|\mathfrak{m}}-\bfL_\mathfrak{m}\tilde{\bfS}^{xx}_{\mathfrak{m}+1|\mathfrak{m}},\ \bfL_\mathfrak{m}\bfS^Q_{\mathfrak{m}},\ \bfL_\mathfrak{m}\bfS_{\mathfrak{m}+1|k}^{xx}\right]\right).\label{eq:auxPsF}
\end{align}
Further details on derivation of \eqref{eq:auxPsF} are given in Appendix~B.



\section{SIR-based Estimator Enhancement}
The above introduced SIR and SIR-based estimators can be seen as \textit{baseline} algorithms, which can be improved to be less computationally demanding and more computationally robust. These improvements are discussed in this section.

\subsection{Efficient Calculation}
In the basic SIR-based algorithms, each moment is calculated using dedicated SIR (Algorithm II). This sequential calculation brings a computational overhead especially in the \textit{filtering} step, where three different measurement-related moments \eqref{eq:LFzpB}--\eqref{eq:LFPzB} are calculated. The computational overhead is caused by the necessity to generate the random orthogonal matrix $\bfC$ and the IR points $\{\bfxi^{(i)}\}_{i=0}^{2n_x}$ at each iteration. To \textit{reduce} the computational complexity, all the moments can be calculated using evaluation of the single SIR, i.e., using the same IR points.

Following this idea, we can calculate the measurement prediction \textit{raw} moments the SIR, i.e.,
\begin{align}
		\hbfz_{k|k-1}&\!\!=\!\!\int\!\!\bfh_{k}(\bfx_{k})\calN\{\bfx_{k};\hbfx_{k|k-1}, \bfP^{xx}_{k|k-1}\}d\bfx_{k},\label{eq:LFzpB1}\\     
		\bfM^{zz}_{k|k-1}&\!\!=\!\!\int\bfh_{k}(\bfx_{k})(\cdot)^T\calN\{\bfx_{k};\hbfx_{k|k-1}, \bfP^{xx}_{k|k-1}\}d\bfx_{k}, \label{eq:LFPzB1}\\
		\bfM^{xz}_{k|k-1}&\!\!=\!\!\int\bfx_{k}(\bfh_{k}(\bfx_{k}))^T\calN\{\bfx_{k};\hbfx_{k|k-1}, \bfP^{xx}_{k|k-1}\}d\bfx_{k}, \label{eq:LFPxzB1}
\end{align}
with one run of the SIR, i.e., using the same weighted IR points, and then, the required \textit{central} moments are 
\begin{align}
    \bfP^{zz}_{k|k-1}&=\bfM^{zz}_{k|k-1}-\hbfz_{k|k-1}\hbfz_{k|k-1}^T+\bfR_k,\label{eq:LFzpBenda}\\
    \bfP^{xz}_{k|k-1}&=\bfM^{xz}_{k|k-1}-\hbfx_{k|k-1}\hbfz_{k|k-1}^T.\label{eq:LFzpBend}
\end{align}

Similarly, the state prediction \eqref{eq:LPxpB}, \eqref{eq:LPPpB} can be implemented in this efficient way, where the cross-covariance matrix $\bfP_{\mathfrak{m},\mathfrak{m}+1|\mathfrak{m}}^{xx}$ \eqref{eq:rstsm_Pxx} can be efficiently calculated as well. This covariance matrix is redundant for the prediction step, but it can be readily used in the smoothing (if smoothing is intended).

If the \textit{square-root} implementation is concerned, analogous modification of the moment calculation can be easily done. Moreover, the filtering and predictive IR points needed for construction of the matrices ${\tbfS}^{xx}_{\mathfrak{m}|\mathfrak{m}}, \tilde{\bfS}^{xx}_{\mathfrak{m}+1|\mathfrak{m}}$ \eqref{eq:predtSxxSM}, \eqref{eq:predtSxxpSM} can be stored during the forward run of the SIF to reduce the smoother computational complexity, but at the expense of the higher memory requirements (depending on the smoothing horizon $\mathfrak{m}$).

\subsection{Integration Error}
SIR implicitly provides information about accuracy, i.e., about the error, of the integral solution. This information is accumulated in $\Sigma_{N}$ \eqref{eq:int_sig_SIR3}. The error assessment can be used either for
\begin{itemize}
    \item Stopping condition of the SIR, which ensures that the calculated moments meet the required accuracy defined by the threshold $\varepsilon_\mathrm{min}$ (indicated in Step 3 of Algorithm II), 
    \item Increase of the state estimate covariance matrix, which then takes into account the SIR output error.
\end{itemize}
The latter option increases the estimate covariance matrix with the mean error assessment. To illustrate this concept in the measurement prediction \eqref{eq:LFzpB1}--\eqref{eq:LFzpBend}, we can calculate the measurement prediction $\hbfz_{k|k-1}$ \eqref{eq:LFzpB1} and the respective error assessment $\Sigma^{\hbfz_{k|k-1}}_{N}$ using the SIR (Algorithm 2) then the predictive covariance matrix can be calculated as
\begin{align}
    \bfP^{zz}_{k|k-1}&=\bfM^{zz}_{k|k-1}-\hbfz_{k|k-1}\hbfz_{k|k-1}^T+\Sigma^{\hbfz_{k|k-1}}_{N}+\bfR_k
\end{align}
instead of \eqref{eq:LFzpBenda}. Utilisation of error assessment of the covariance matrices calculation, i.e., of $\Sigma^{\bfP_{k|k-1}^{zz}}_{N}$ and $\Sigma^{\bfP_{k|k-1}^{xz}}_{N}$ is currently under investigation.

\subsection{Integration Rule Order}
We have introduced the SIR of the 3-rd order. However, the SIR can be designed for any odd degree (typically, first and fifth), which offers the trade-off between the computational complexity and integration error \cite{DuStSiBl:15,GeMo:98}.

\section{Stone Soup and Implementation}

The Stone Soup project project is an open-source tracking and estimation framework currently available as a python library; to support developers, users, and systems engineers. The framework saw its first public alpha release in 2017 \cite{ThomasPaulA2017Aosf} with the first beta release in 2019. The name of the framework is inspired by the tale of the “Stone Soup”; from European folklore in which many ingredients (estimation methods) are contributed by villagers (researchers) to devise a flavourful soup(useful software). Inherent in the common repository is the ability to compare and reuse contributions to extend the quality of the estimation capabilities. While some have sought initial ideas \cite{BlaschEruj2020prc, ORourkeSeanM2021IoEK}, Barr et al. \cite{BarrJordi2022SSos} highlight the many capabilities such as track generation, filtering, classification, data association, and sensor management routines. Examples are provided for video tracking, automatic dependent surveillance – broadcast (ADS-B) tracking, orbital estimation, drone tracking, sensor management, and tracking evaluation. Recently, a series of papers began using the techniques, specifically for radar-based tracking. Carniglia et al. \cite{BalajiBhashyam2022IoSB} used the Stone Soup Generalized Optimal Sub-Pattern Assignment (GOSPA) and Single Integrated Air Picture (SIAP) metrics to compare the Joint-Probabilistic Data Association (JPDA) to track multiple airborne targets being in clutter from two ground-based radars to assess the track quality sensitivity of biased sensor measurements. Similarly, tracking of the adaptive Kernel Kalman Filter (AKKF) was compared to the particle filter (PF) using Stone Soup infrastructure \cite{WrightJames0256739}.

Given the recent interest in artificial intelligence techniques, Stone Soup was used to compare neural net (NN) trackers for drone surveillance using the recurrent NN (RNN), convolutional NN (CNN), and machine learning perceptron (MLP) networks \cite{GoodallFinn10149795} as well as drone tracking with reinforcement learning. The current techniques of Stone Soup are summarised in \cite{AlhadhramiEsra2023DTAU} with models (measurement, transition), filters (extended or unscented KF, PF), state update methods (information, Gromov Flow and PHD), and simulation capabilities (metrics, scenario generators, and track stitching). Among the many capabilities of Stone Soup, it provides a useful infrastructure for novel tracking and estimation method comparisons.

\begin{figure}
    \centering
    \includegraphics[width=0.95\linewidth]{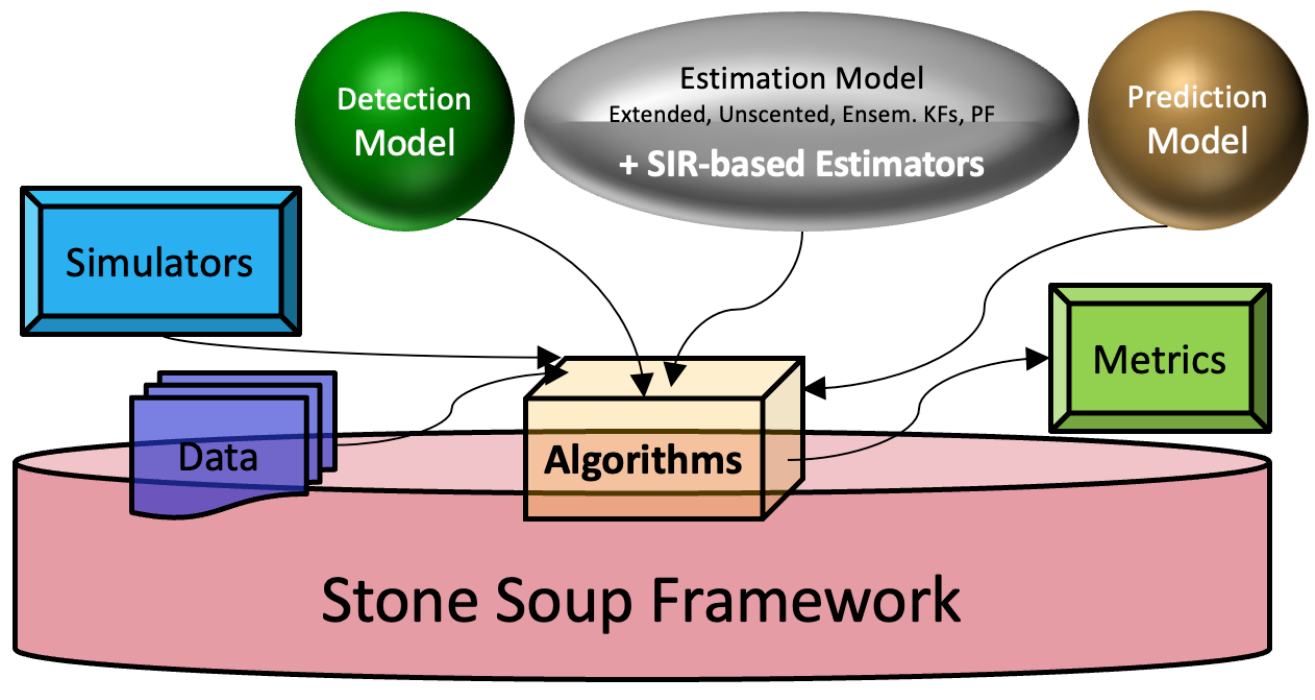}
    \caption{Elements of the ISIF Stone Soup Framework with SIR-based estimators \cite{BlaschEruj2020prc}.}
    \label{fig:stonesoup}
    \vspace*{-7mm}
\end{figure}

In the spirit of collaboration, our group is providing the SIF and an associated smoother and predictor for inclusion in the framework. Integration in Stone Soup also allows for more easy comparisons between the SIR-based estimators and other estimation methods. Of particular interest here, is the relationship between the unscented Kalman filter (UKF) and the SIF, as the two algorithms are the most similar algorithms in the framework. In fact, the UKF can be seen as a special case of the SIF, when the IR (or sigma) points are suitably generated \cite{DuStSiBl:15}. The overall scheme of the Stone Soup framework with the SIR-based estimators is illustrated in Figure \ref{fig:stonesoup}.

\subsection{Stochastic Integration Filter, Smoother, and Predictor Implementation: Current Status}
Currently, we have implemented the stochastic integration rule of the 1st, 3rd, and 5th order for Gaussian weighted integral solution and applied it in the SIF predictor, SIF updater, and SIF smoother, introduced and derived in this paper. These algorithms are submitted to be included in the Stone Soup repository. The square-root versions are in development and will be submitted shortly after the publication of this paper.


\section{Numerical Illustration}
Performance of the SIF is evaluated using the tracking scenario used for illustration of the extended KF (EKF) and the UKF \cite{AnMo:79,Si:06,JuUhl:04,DuSiStr:12} in Stone Soup\footnote{\url{https://stonesoup.readthedocs.io/en/v1.1/auto_tutorials/index.html}}. In particular, the following state-space model is considered
\begin{align}
  \bfx_{k+1}&=\left[\begin{smallmatrix}
     1 & T & 0 & 0 \\ 0 & 1 & 0 & 0 \\
    0 & 0 & 1 & T \\ 0 & 0 & 0 & 1\end{smallmatrix}\right]\bfx_k+\bfw_k, \label{eq:motion}\\
 \bfz_{k}&=\left[\begin{smallmatrix}
     \arctan\frac{x_{3,k} - r_y}{x_{1,k} -r_x}\\\sqrt{(x_{1,k} - r_x)^2+(x_{3,k} -r_y)^2}
     \end{smallmatrix}\right]+\bfv_k, \label{eq:measurement}
\end{align}
where the state is represented by an sought object position and velocity in $x$ and $y$ directions and the measurement is the bearing angle\footnote{Bearing measurement model is subject to wrapping.} and range of the object from a radar located at $\bfr=[r_x,r_y]^T=[50, 0]^T$. Sampling period is $T=1$ and  $k = 0, 1,\ldots, 20$. The state and measurement noises covariance matrices \eqref{eq:pw}, \eqref{eq:pv} are
\begin{align}
    \bfQ_k=\left[\begin{smallmatrix}
     q_1\tfrac{T^3}{3} & q_1\tfrac{T^2}{2} & 0 & 0 \\ q_1\tfrac{T^2}{2} & q_1T & 0 & 0 \\
    0 & 0 & q_2\tfrac{T^3}{3} & q_2\tfrac{T^2}{2}\\ 0 & 0 & q_2\tfrac{T^2}{2} & q_2T\end{smallmatrix}\right], \bfR_k=\left[\begin{smallmatrix}
        0.2\pi/180 & 0 \\ 0 & 1
    \end{smallmatrix}\right]
\end{align}
with $q_1=q_2=0.05$. The initial state \eqref{eq:px0} description is
\begin{align}
    \bbfx_0=\left[\begin{smallmatrix}
     50  \\ 1  \\ 1 \\ 1\end{smallmatrix}\right], \bfP_0=\left[\begin{smallmatrix}
     1.5 & 0 & 0 & 0 \\ 0 & 0.5 & 0 & 0 \\
    0 & 0 & 1.5 & 0 \\ 0 & 0 & 0 & 0.5\end{smallmatrix}\right].\label{eq:motion_init}
\end{align}

The estimation performance is assessed using
\begin{itemize}
    \item The root-mean-square error (RMSE) of the $i$-th state
    \begin{align}
        \mathrm{RMSE}_{i}=\sqrt{\tfrac{1}{T}\sum_{k=0}^T(x_{i,k}-\hat{x}_{i,k|k})^2},
    \end{align}
    \item The averaged normalised estimation error squared (ANEES)
    \begin{align}
        \mathrm{ANEES}=\tfrac{1}{T}\sum_{k=0}^T(\bfx_{k}-\hat{\bfx}_{k|k})^T({\bfP}^{xx}_{k|k})^{-1}(\cdot),
    \end{align}
\end{itemize}
calculated over $10^4$ Monte-Carlo (MC) simulations, where $\hat{x}_{i,k|k}$ is the filtering estimate of the $i$-th element of the state $x_{i,k}$. The UKF was designed with the default scaling parameters of the Stone Soup implementation, i.e., with $\alpha=0.5, \beta=2, \kappa=3-n_x=-1$ and the SIF with maximum number of iterations $N_{\max}=10$.


\begin{table}
\centering
\begin{tabular}{ll|ccc|c}
 &  & EKF & UKF & \textbf{SIF} & SIF Improvement \\ 
\hline 
\multirow{4}{*}{RMSE} &  $x_1$ & 0.7774 & 0.8344 & \textbf{0.7398} & \multirow{4}{*}{5\%} \\ 
  & $x_2$ &  0.3867 & 0.3987  & \textbf{0.3881} & \\ 
    & $x_3$ & 0.6884 & 0.7343  & \textbf{0.6781} &\\ 
    & $x_4$ & 0.3758 &  0.3808  & \textbf{0.3732} & \\ 
\hline 
  ANEES  &  & 10.9559 & 5.3510  & \textbf{4.0810} & 40\%\\ 
\hline 
\end{tabular}\caption{EKF, UKF, and SIF performance summary.}\label{tab:res}
\vspace*{-6mm}
\end{table}

Excellent performance of the SIF was demonstrated in various scenarios and under different settings \cite{DuStrSi:13,DuStSiBl:15,KhReAb:17,ZhGu:17,StDu:17}. This is also confirmed using the Stone Soup defined tracking scenario \eqref{eq:motion}--\eqref{eq:motion_init} run using the Python and MATLAB\footnote{MATLAB\textregistered\ implementation of the unscented and SIR-based estimators is available at \url{https://github.com/IDM-UWB/FUSION2024_SIF}.}\textregistered\ implementations and evaluated using the ANEES and RMSE metrics. The results are shown in Table~\ref{tab:res}, where the absolute values and an \textit{average} improvement of the SIF w.r.t. EKF and UKF can be found. In terms of the RMSE all the considered filters provide very similar performance\footnote{The EKF estimate diverged in a small number of the MC simulations.}, however, as indicated by the ANEES, the UKF and \textit{especially} the SIF provide more consistent estimates (i.e., more realistic estimate covariance matrix).  



Regarding the computational complexity, the SIF is more demanding than both considered GEs as it processes high(er) number of IR points. However, the computational complexity significantly depends not only on the implementation, but also by the selected programming language, environment, and target platform. For this reason and w.r.t. to publicly available implementations in Python and MATLAB\textregistered, we would leave evaluation on the user.

\section{Concluding Remarks}
This paper dealt with state estimation of the nonlinear models using the Gaussian estimators. The emphasis was laid on the evaluation of the Gaussian weighted integrals needed for the state estimate moment calculation using the stochastic integration rule. Compared to other solutions to the integrals, the SIR provides asymptotically unbiased moment estimates. The SIR-based filter, denoted as the \textit{stochastic integration filter}, was complemented with the multi-step \textit{predictor} and \textit{smoother} in the \textit{full} and \textit{square-root} forms. The SIR-based estimators have been implemented in Python for the Stone Soup infrastructure and in MATLAB\textregistered\ for convenience. Both implementations provide consistent results. The SIR-based estimators offer improved estimation performance over the extended and unscented KFs in a Stone Soup defined tracking scenario.

\vspace{0.2em}
 \textit{Acknowledgement}: J. Duník, J. Matoušek, and O.~Straka's work was co-funded by the European Union under the project ROBOPROX - Robotics and advanced industrial production  (reg. no. CZ.02.01.01\slash00\slash22\_008\slash0004590). J.~Hiles and R. Niu's work was supported in part under FA9550-22-1-0038. The views and conclusions contained in this paper are those of the authors and should not be interpreted as representing the official policies, either expressed or implied, of the U.S. Government. 

\section*{Appendix A}
The sketch of the proof of \eqref{eq:allPoints}, \eqref{eq:allWeights} stems from the SIR (Algorithm II) and a recursive calculation of the integral $\hat{\mathcal{I}}_N$ \eqref{eq:int_val_SIR3}, which can be written for increasing $N$ from 1 to 3 (assuming $\bfg(\cdot)$ for simplicity to be scalar function of vector argument) as
\begin{align}
    \hat{\mathcal{I}}_1 &= \mathfrak{I}^{(1)}=\sum_{i=0}^{2n_x}\bfg(\bfxi_{q|l}^{(1,i)})\omega_{q|l}^{(1,i)},\\
    \hat{\mathcal{I}}_2 &= (1-\tfrac{1}{2})\hat{\mathcal{I}}_1+\mathfrak{I}^{(2)}/2=\mathfrak{I}^{(1)}/2+\mathfrak{I}^{(2)}/2,\\
    \hat{\mathcal{I}}_3 &= (1-\tfrac{1}{3})\hat{\mathcal{I}}_2+\mathfrak{I}^{(3)}/3=\tfrac{2}{3}(\mathfrak{I}^{(1)}/2+\mathfrak{I}^{(2)}/2)+\mathfrak{I}^{(3)}/3\nonumber\\
    &=\mathfrak{I}^{(1)}/3+\mathfrak{I}^{(2)}/3+\mathfrak{I}^{(3)}/3,
\end{align}
where the $i$-term can be written 
\begin{align}
    \tfrac{\mathfrak{I}^{(n)}}{N}\!=\!\tfrac{1}{N}\sum_{i=0}^{2n_x}\bfg(\bfxi_{q|l}^{(n,i)})\omega_{q|l}^{(n,i)}\!\!=\!\!\sum_{i=0}^{2n_x}\bfg(\bfxi_{q|l}^{(n,i)})\tilde{\omega}_{q|l}^{(n,i)},
\end{align}
which directly leads to \eqref{eq:allPoints}, \eqref{eq:allWeights}.

\section*{Appendix B: Matrix Factorisation}
\subsection{Filtering Covariance Matrix}
To show identity of the filtering covariance matrix forms \eqref{eq:LFPfB} and \eqref{eq:LFPfBFF}, let us start with multiplication and expression of the right-hand side of \eqref{eq:LFPfBFF} 
\begin{align}
    \bfP^{xx}_{k|k}&= \tbfS^{xx}_{k|k-1}(\tbfS^{xx}_{k|k-1})^T-\tbfS^{xx}_{k|k-1}(\bfK_k\tilde{\bfS}^{zz}_{k|k-1})^T\label{eq:aux1}\\
    &-\bfK_k\tilde{\bfS}^{zz}_{k|k-1}(\tbfS^{xx}_{k|k-1})^T+\bfK_k\tilde{\bfS}^{zz}_{k|k-1}(\cdot)^T+\bfK_k\bfS^R_{k}(\cdot)^T\nonumber
\end{align}
and utilisation of the following identities \cite{SiDu:05}
\begin{align}
 &\bfK_k\bfP^{zz}_{k|k-1}\bfK^T_k = \bfP^{xz}_{k|k-1}\bfK^{T}_k = \bfK_k(\bfP^{xz}_{k|k-1})^T\label{eq:fr0}\\
 &=\tilde{\bfS}^{xx}_{k|k-1}(\tilde{\bfS}^{zz}_{k|k-1})^T\bfK^T_k=\bfK_k\tilde{\bfS}^{zz}_{k|k-1}(\tilde{\bfS}^{xx}_{k|k-1})^T,\\
 &= \bfK_k\left(\tilde{\bfS}^{zz}_{k|k-1}(\tilde{\bfS}^{zz}_{k|k-1})^T+\bfS^{R}_{k}(\bfS^{R}_{k})^T\right)\bfK^T_k.\label{eq:fr3}
\end{align}
Substitution of \eqref{eq:fr0}--\eqref{eq:fr3} into \eqref{eq:aux1} and w.r.t. Algorithm II and \eqref{eq:allPoints}, \eqref{eq:allWeights}, the covariance matrix \eqref{eq:LFPfBFF} reads
\begin{align}
     \bfP^{xx}_{k|k}&=\tbfS^{xx}_{k|k-1}(\tbfS^{xx}_{k|k-1})^T-\bfK_k\bfP^{zz}_{k|k-1}\bfK^T_k\nonumber\\
    &-\bfK_k\bfP^{zz}_{k|k-1}\bfK^T_k+\bfK_k\bfP^{zz}_{k|k-1}\bfK^T_k,\label{eq:PxxF}
\end{align}
which is identical with \eqref{eq:LFPfB}.

\subsection{Smoothing Covariance Matrix}
Multiplication of \eqref{eq:auxPs} right-hand side leads
\begin{align}
    &\bfP_{\mathfrak{m}|k}^{xx}\!=\!\tbfS^{xx}_{\mathfrak{m}|\mathfrak{m}}(\tbfS^{xx}_{\mathfrak{m}|\mathfrak{m}})^T\!-\!\tbfS^{xx}_{\mathfrak{m}|\mathfrak{m}}(\bfL_\mathfrak{m}\tilde{\bfS}^{xx}_{\mathfrak{m}+1|\mathfrak{m}})^T\!+\!\bfL_\mathfrak{m}\bfS^Q_{\mathfrak{m}}(\cdot)^T\nonumber\\
    &\!-\!\bfL_\mathfrak{m}\tilde{\bfS}^{xx}_{\mathfrak{m}+1|\mathfrak{m}}(\tbfS^{xx}_{\mathfrak{m}|\mathfrak{m}})^T\!+\!\bfL_\mathfrak{m}\tilde{\bfS}^{xx}_{\mathfrak{m}+1|\mathfrak{m}}(\cdot)^T\!+\!\bfL_\mathfrak{m}\bfS_{\mathfrak{m}+1|k}^{xx}(\cdot)^T.\label{eq:PsFaux}
\end{align}
Substituting  the following identities \cite{SiDu:06} 
\begin{align}
 &\bfL_{\mathfrak{m}}\bfP_{\mathfrak{m}+1|\mathfrak{m}}^{xx}\bfL^T_\mathfrak{m} = \bfP_{\mathfrak{m},\mathfrak{m}+1|\mathfrak{m}}^{xx}\bfL_{\mathfrak{m}}^T = \bfL_{\mathfrak{m}}(\bfP_{\mathfrak{m},\mathfrak{m}+1|\mathfrak{m}}^{xx})^T\nonumber\\
 &=\tilde{\bfS}^{xx}_{\mathfrak{m}|\mathfrak{m}}(\tilde{\bfS}^{xx}_{\mathfrak{m}+1|\mathfrak{m}})^T\bfL^T_\mathfrak{m}=\bfL_\mathfrak{m}\tilde{\bfS}^{xx}_{\mathfrak{m}+1|\mathfrak{m}}(\tilde{\bfS}^{xx}_{\mathfrak{m}|\mathfrak{m}})^T,\\
 &= \bfL_\mathfrak{m}\left(\tilde{\bfS}^{xx}_{\mathfrak{m}+1|\mathfrak{m}}(\tilde{\bfS}^{xx}_{\mathfrak{m}+1|\mathfrak{m}})^T+\bfS^{Q}_{\mathfrak{m}}(\bfS^{Q}_{\mathfrak{m}})^T\right)\bfL^T_\mathfrak{m},\label{eq:fr6}
\end{align}
into \eqref{eq:PsFaux}. W.r.t. Algorithm II, \eqref{eq:rstsm_x}--\eqref{eq:rstsm_Pxx}, \eqref{eq:allPoints}, \eqref{eq:allWeights}, we get
\begin{align}
   \bfP_{\mathfrak{m}|k}^{xx}&=\bfP_{\mathfrak{m}|\mathfrak{m}}^{xx}+\bfL_{\mathfrak{m}}\bfP_{\mathfrak{m}+1|k}^{xx}\bfL_{\mathfrak{m}}^T-\bfL_{\mathfrak{m}}\bfP_{\mathfrak{m}+1|\mathfrak{m}}^{xx}\bfL_{\mathfrak{m}}^T\nonumber\\
   &+\bfL_{\mathfrak{m}}\bfP_{\mathfrak{m}+1|\mathfrak{m}}^{xx}\bfL_{\mathfrak{m}}^T-\bfL_{\mathfrak{m}}\bfP_{\mathfrak{m}+1|\mathfrak{m}}^{xx}\bfL_{\mathfrak{m}}^T,\label{eq:Psm_ext}
\end{align}
which is identical with \eqref{eq:rstsm_P}.

	%
	%
	\bibliographystyle{splncs_srt}


\begin{thebibliography}{10}
		
		\bibitem{AlhadhramiEsra2023DTAU}
		Alhadhrami, E., Seghrouchni, A.E.F., Barbaresco, F., Zitar, R.A.:
		\newblock Drones tracking adaptation using reinforcement learning: Proximal
		policy optimization.
		\newblock In: 2023 24th International Radar Symposium (IRS). Volume 2023 24th
		International Radar Symposium (IRS)., German Institute of Navigation (DGON)
		(2023)  1--10
		
		\bibitem{AnMo:79}
		Anderson, B.D.O., Moore, J.B.:
		\newblock Optimal Filtering.
		\newblock Prentice Hall, New Jersey (1979)
		
		\bibitem{ArHa:09}
		Arasaratnam, I., Haykin, S.:
		\newblock Cubature {K}alman filters.
		\newblock IEEE Trans. on Automatic Control \textbf{54}(6) (2009)  1254--1269
		
		\bibitem{Arn:04}
		Arndt, C.:
		\newblock Information measures.
		\newblock Springer (2004)
		
		\bibitem{BarrJordi2022SSos}
		Barr, J., Harrald, O., Hiscocks, S., Perree, N., Pritchett, H., Vidal, S.,
		Wright, J., Carniglia, P., Hunter, E., Kirkland, D., Raval, D., Zheng, S.,
		Young, A., Balaji, B., Maskell, S., Hernandez, M., Vladimirov, L.:
		\newblock {Stone Soup} open source framework for tracking and state estimation:
		enhancements and applications.
		\newblock Volume 12122., SPIE (2022)  1212205--1212205--17
		
		\bibitem{BlaschEruj2020prc}
		Blasch, E., Niu, R., O'Rourke, S.:
		\newblock Target tracking analysis for {Stone Soup}.
		\newblock In: Int'l Conf. on Information Fusion. (2020)
		
		\bibitem{BalajiBhashyam2022IoSB}
		Carniglia, P., Balaji, B., Damini, A.:
		\newblock Investigation of sensor bias and signal quality on target tracking
		with multiple radars.
		\newblock In: Int'l Inst. and Measurement Tech. Conf. (2022)
		
		\bibitem{DuStrSi:13a}
		Dun\'{i}k, J., Straka, O., \v{S}imandl, M.:
		\newblock Nonlinearity and non-{G}aussianity measures for stochastic dynamic
		systems.
		\newblock In: Proceedings of the 16th International Conference on Information
		Fusion, Istanbul (2013)
		
		\bibitem{DuStrSi:13}
		Dun\'{i}k, J., Straka, O., \v{S}imandl, M.:
		\newblock Stochastic integration filter.
		\newblock IEEE Transactions on Automatic Control \textbf{58}(6) (2013)
		1561--1566
		
		\bibitem{DuStSiBl:15}
		Dun\'{i}k, J., Straka, O., \v{S}imandl, M., Blasch, E.:
		\newblock Random-point-based filters: Analysis and comparison in target
		tracking.
		\newblock IEEE Transactions on Aerospace and Electronic Systems \textbf{51}(2)
		(2015)  303--308
		
		\bibitem{DuSiStr:12}
		Dun\'{i}k, J., \v{S}imandl, M., Straka, O.:
		\newblock Unscented {K}alman filter: Aspects and adaptive setting of scaling
		parameter.
		\newblock IEEE Transactions on Automatic Control \textbf{57}(9) (2012)
		2411--2416
		
		\bibitem{GeMo:98}
		Genz, A., Monahan, J.:
		\newblock Stochastic integration rules for infinite regions.
		\newblock SIAM Journal on Scientific Computing \textbf{19}(2) (1998)  426--439
		
		\bibitem{GoodallFinn10149795}
		Goodall, F., Ahmad, B.I.:
		\newblock Adaptation of multi-target tracker using neural networks in drone
		surveillance radar.
		\newblock In: 2023 IEEE Radar Conference (RadarConf23). (2023)  1--6
		
		\bibitem{ORourkeSeanM2021IoEK}
		Hiles, J., O'Rourke, S.M., Niu, R., Blasch, E.P.:
		\newblock Implementation of ensemble kalman filters in stone-soup.
		\newblock In: Int'l Conf. on Information Fusion. (2021)
		
		\bibitem{JuUhl:04}
		Julier, S.J., Uhlmann, J.K.:
		\newblock Unscented filtering and nonlinear estimation.
		\newblock IEEE Proceedings \textbf{92}(3) (2004)  401--421
		
		\bibitem{KhReAb:17}
		Khalid, S.S., Rehman, N.U., Abrar, S.:
		\newblock Robust stochastic integration filtering for nonlinear systems under
		multivariate t-distributed uncertainties.
		\newblock Signal Processing \textbf{140} (2017)  53--59
		
		\bibitem{NoPoRa:00b}
		N{\o}rgaard, M., Poulsen, N.K., Ravn, O.:
		\newblock New developments in state estimation for nonlinear systems.
		\newblock Automatica \textbf{36}(11) (2000)  1627--1638
		
		\bibitem{SaOrAcAc:21}
		Santos-León, J.C., Orive, R., Acosta, D., Acosta, L.:
		\newblock The cubature {K}alman filter revisited.
		\newblock Automatica \textbf{127} (2021)  109541
		
		\bibitem{Sa:13}
		S\"{a}rkk\"{a}, S.:
		\newblock Bayesian Filtering and Smoothing.
		\newblock Cambridge University Press (2013)
		
		\bibitem{Si:06}
		Simon, D.:
		\newblock Optimal State Estimation: {K}alman, H Infinity, and Nonlinear
		Approaches.
		\newblock Wiley-Interscience (2006)
		
		\bibitem{So:74}
		Sorenson, H.W.:
		\newblock On the development of practical nonlinear filters.
		\newblock Information Sciences \textbf{7} (1974)  230--270
		
		\bibitem{StDu:17}
		Straka, O., Duník, J.:
		\newblock Stochastic integration student's-t filter.
		\newblock In: 2017 20th International Conference on Information Fusion
		(Fusion), Xi'an, China (2017)
		
		\bibitem{ThomasPaulA2017Aosf}
		Thomas, P.A., Barr, J., Balaji, B., White, K.:
		\newblock An open source framework for tracking and state estimation ('stone
		soup').
		\newblock In: SPIE Proceedings. Volume 10200., SPIE (2017)
		1020008--1020008--10
		
		\bibitem{SiDu:05}
		\v{S}imandl, M., Dun\'{i}k, J.:
		\newblock Sigma point gaussian sum filter design using square root unscented
		filters, volume 1.
		\newblock In: Proceedings of the 16th IFAC World Congress, Prague, Czech
		Republic, Oxford: Elsevier (2005)
		
		\bibitem{SiDu:06}
		\v{S}imandl, M., Dun\'{i}k, J.:
		\newblock Design of derivative-free smoothers and predictors.
		\newblock IFAC Proceedings Volumes \textbf{39}(1) (2006)  1240--1245 14th IFAC
		Symposium on Identification and System Parameter Estimation (SYSID).
		
		\bibitem{SiDu:09}
		\v{S}imandl, M., Dun\'{i}k, J.:
		\newblock Derivative-free estimation methods: New results and performance
		analysis.
		\newblock Automatica \textbf{45}(7) (2009)  1749--1757
		
		\bibitem{WrightJames0256739}
		Wright, J.S., Hopgood, J.R., Davies, M.E., Proudler, I.K., Sun, M.:
		\newblock Implementation of adaptive kernel kalman filter in {Stone Soup}.
		\newblock In: 2023 Sensor Signal Processing for Defence Conference (SSPD).
		(2023)  1--5
		
		\bibitem{Yang:08}
		Yang, C., Blasch, E.:
		\newblock Fusion of tracks with road constraints.
		\newblock J. Adv. Information Fusion \textbf{3} (2008)  14--32
		
		\bibitem{ZhGu:17}
		Zhou, D., Guo, L.:
		\newblock Stochastic integration {H}$\infty$ filter for rapid transfer
		alignment of ins.
		\newblock Sensors \textbf{17}(11) (2017)
		
	\end{thebibliography}

\end{document}